\documentclass{llncs}

\usepackage{color}
\usepackage{tikz}
\usetikzlibrary{arrows,automata,shapes,backgrounds,fit,decorations.pathreplacing}
\usepackage{units,multirow,subfigure,paralist}

\usepackage{amssymb,amsmath}
\usepackage{stmaryrd,wasysym}
\usepackage{xspace}
\usepackage{url}
\usepackage{extarrows}
\usepackage{mathtools}
\usepackage{listings}

\makeatletter
   \def\marginparright{\@mparswitchfalse}
   \def\marginparoutside{\@mparswitchtrue}
\makeatother
\def\comment#1{}

\def\withcomments{
\addtolength{\oddsidemargin}{-0.5 in}
\addtolength{\evensidemargin}{-0.5 in}
\newcounter{mycommentcounter}
\marginparright
\def\comment##1{\refstepcounter{mycommentcounter}%
  \ifhmode%
  \unskip%
  {\dimen1=\baselineskip \divide\dimen1 by 2 %
    \raise\dimen1\llap{\tiny {\bf \color{red}*-\themycommentcounter-*}}}\fi%
  \marginpar{\renewcommand{\baselinestretch}{0.8}%
    \footnotesize [\themycommentcounter]: \raggedright ##1}}
\date{\framebox{Draft of \today}}
}

\newtheorem{Def}{Definition}
\newtheorem{Ex}{Example}

\lstdefinelanguage{lgmpl} {alsoletter={\#, \&},
  mathescape=true,
  basicstyle={\small \tt},
  escapechar=\@,
  boxpos=c,
  morekeywords= {theorem, lemma, apply, by, constdefs, definition, where,
    infixl, types, consts, primrec, have, from, show, let,  proof, qed, is,
    sorry, with, assume, fix, thus, hence, datatype, if, then, else, in, case,
    of
  },
  emph={[2]int, char, string, self, boolean},
  emphstyle={[2]\it},
literate=
  {->}{{$\rightarrow$}}2
  {=>}{{$\Rightarrow$}}2
  {-->}{{$\Rightarrow$}}2
  {\\forall}{{$\forall$}}2
  {\\exist}{{$\exists$}}2
  {\\exists}{{$\exists$}}2
  {AND}{{$\& \&$}}3
  {ALL}{{$\forall$}}2
  {EX}{{$\exists$}}2
  {\%}{{$\lambda$}}1
  {\\\/}{{$\sqcap$}}2
  {|-}{{$\vdash$}}2
  {==>}{{$\Rightarrow$}}2
  {~}{{$\neg$}}1
  {~=}{{$\neq$}}2
  {;}{{\tt ;}}1
  { graph }{{ $G$ }}2
}

\lstnewenvironment{gmpl} {
  \lstset{extendedchars=true,columns=flexible}
  \lstset{language=lgmpl,deletestring=[b]', basicstyle=\scriptsize \tt}%
} {}

\newcommand{\gm}[1]{{\lstinline[language=lgmpl, basicstyle={\footnotesize \tt} ]@#1@}}

\newcommand{\etal}{\emph{et al.}~}

\def\all{\ensuremath{{\mathsf{all}}}}

\def\A{\ensuremath{\mathcal{A}}}

\def\I{\ensuremath{\mathcal{I}}}

\def\P{\ensuremath{\mathcal{P}}}

\def\R{\ensuremath{\mathcal{R}}}
\def\W{\ensuremath{\mathcal{W}}}
\def\U{\ensuremath{\mathcal{U}}}

\def\allow{\ensuremath{\mathbf{allow}}\xspace}
\def\deny{\ensuremath{\mathbf{deny}}\xspace}

\def\unique{\ensuremath{\mathbf{unique}}\xspace}
\def\once{\ensuremath{\mathbf{once}}\xspace}
\def\all{\ensuremath{\mathbf{all}}\xspace}

\begin{document}
    \mainmatter  

    \title{Implementing Access Control Markov Decision Processes with GLPK/GMPL\thanks{This work is supported by the EU project NESSoS.}}

    \titlerunning{}

    %
    %
    \author{Charles Morisset}
    
    \authorrunning{}

    
    \institute{IIT-CNR, Security Group\\
      Via Giuseppe Moruzzi 1, 56124 Pisa, Italy, \\
     {\tt charles.morisset@iit.cnr.it}
    }

    \maketitle

\begin{abstract}
In a recent approach, we proposed to model an access control mechanism as a Markov Decision Process, 
thus claiming that in order to make an access control decision, one can use well-defined mechanisms from 
decision theory. We present in this paper an implementation of such mechanism, using the open-source 
solver GLPK, and we model the problem in the GMPL language. We illustrate our approach with a simple, yet 
expressive example, and we show how the variation of some parameters can change the final outcome. In 
particular, we show that in addition to returning a decision, we can also calculate the value of each decision. 
\end{abstract}

\section{Introduction}
\label{sec:introduction}

An access control mechanism is responsible within an information system of intercepting any access made by a user
over a resource, deciding whether this access should be allowed or not, and enforcing the corresponding decision. Most existing 
mechanisms rely on the definition of an {\em access control policy}, defined in a dedicated language (e.g., RBAC~\cite{Ferraiolo99}, XACML~\cite{XACML}), 
which can be roughly seen as a set of rules separating the set of accesses into the secure ones and the non-secure ones.

We recently proposed in~\cite{MM12} to model an access control mechanism as a Markov Decision Process (MDP)~\cite{Bellman57}. The main strength of this approach 
is to account for uncertainty and probabilistic behaviour of the system directly into the decision mechanism. From a global point of view, such 
a modelling expresses the fact that making an access control decision can be done using tools from decision theory. 
By using an MDP instead of a traditional security mechanism, one can express an intuitive
notion of {\em utility} rather than a standard classification between secure and non-secure situations. 
This is particularly useful in some critical environments, where at a given point in time, it might only be possible to choose between ``bad'' situations. 

For instance, a study~\cite{Rostad:2006:SAC:1191820.1191875} revealed that in a particular healthcare system, 
74\% of users have manual overriding permissions and 54\% of records have been accessed at least once using an overriding permission. 
Such situations are due to the conflicting choice between respecting the static policy and providing the best possible care to the patients. 
The typical example is that of a nurse, who is normally not authorized to access the medical record of a patient, and who needs to access it in order to deliver some treatment
while no attending physician is available. The information system can only choose between two ``bad'' options: granting the access to the nurse, 
and therefore breaching patient confidentiality, or denying the access, and therefore risking the life of the patient.

We present in this paper a detailed example of the implementation of an Access Control-Markov Decision Process (AC-MDP) using GLPK/GMPL. The objective of this paper 
is therefore to present a basic and simple guide of what to do and what to expect when implementing an AC-MDP. In order to do so, we introduce a simple running example, that we 
will tweak along the paper in order to illustrate how changing some parameters might change the results. We try to present our approach from the perspective of a security engineer responsible 
for implementing a security mechanism for a concrete problem, loosely inspired from the healthcare context. The rest of this paper is therefore organized as follows: in Section~\ref{sec:problem}, 
we state the concrete problem we want to implement and we recall the basic definition of an AC-MDP. In Section~\ref{sec:general}, we present the general implementation in GLPK/GMPL of the AC-MDP for this concrete
problem, and we instantiate this general model in Section~\ref{sec:concrete}. Finally, we discuss the different results and the important points to extract from our approach in Section~\ref{sec:discussion}. 

\subsection*{Related Work}
To the best of our knowledge, our recent approach~\cite{MM12} was the first usage of Markov Decision Processes in the context of access control systems. However, the problem of dealing with risk and uncertainty for access control systems has been already studied in the literature. For instance, Aziz \etal \cite{DBLP:journals/jhsn/AzizFHS06} refine a policy to a more restrictive one, in order to deal with threats. 

Risk is often considered as an input to the system, that must stay below a certain threshold. 
Cheng \etal introduce in~\cite{CHEN-07-SP} the Fuzzy Multi-Level Security model, where each access is associated with a level of risk, and the final decision of the authorization mechanism is given according to some predefined risk thresholds. Diep \etal extend this approach in ~\cite{Diep:2007:EAC:1251989.1253389} by considering costs in terms of availability, integrity and confidentiality for each decision, and use thresholds for each corresponding risk. 

Some approaches aim at calculating the risk from the environment. For instance, Ni \etal introduce in~\cite{NI-10-ASIACCS} fuzzy security parameters that can be inferred from traditional parameters, hence introducing a notion of uncertainty directly into the parameters of the system, while Chen and Crampton present in~\cite{chen:stm11} a way to calculate the risk for a RBAC model, using intuitive notions of competence and distance between users, roles and permissions. The probabilistic change of security attributes is considered by Krautsevich
\etal in~ \cite{Krautsevich10Risk-aware}, who model this change using a Markov chain. 

Dealing with uncertainty requires quantitative techniques, and therefore accesses must be associated with a utility value. Some models use these notions, for instance Krautsevich \etal extend in~\cite{KMMY11} the auto-delegation mechanism~\cite{cram:stm10} with probabilistic availability, 
reusing some notions of utility functions previously introduced in~\cite{Krautsevich10Influence}. 
Similarly, Molloy \etal present in~\cite{MDMCLR11} a model to predict and make local decisions under uncertainty, where the system needs to choose between taking a decision locally or defer it to a central server, according to the utility of the access and the cost of communicating with the server.

\medskip

Beyond the scope of access control, several pieces of work use the concept of Markov Decision Process (MDP)
in the context of security. For instance, Kreidl
introduces in~\cite{Kreidl:2010:AMD:1908637.1909503}  a simple MDP with only three states (normal, under attack and failure) and three decisions (wait, defend and reset), which analyses the cost of defending countermeasures against the cost of an intrusion. 

Similarly, He \etal present in~\cite{5365272} an analysis of the operational costs and the negative and positive impact of security countermeasures using Domain Partitional Markov Decision Processes, which partition the network into several security domains, each domain coming with its own MDP. This work, as the previous one, mostly focuses on the detection of intrusions and the decisions needed when some are discovered. 

Finally, Singh \etal use in~\cite{Singh:2010:MDP:1741400.1741462} an MDP to make channel assignments for network devices. This problem can be considered as a special instance of an access control problem, where the devices ask to access channels, with a specific policy stating that any device can access a channel, but the more devices use a given channel, the lower is the utility for this channel. Hence, our approach can be seen as a more general approach, where the policy is not constrained. 

\section{Problem Statement}
\label{sec:problem}

\def\alice{\textbf{alice}\xspace}
\def\bob{\textbf{bob}\xspace}
\def\low{\textbf{low}\xspace}
\def\high{\textbf{high}\xspace}

As we said in the Introduction, we present our approach from the perspective of security engineer, who is responsible for implementing a security mechanism for a concrete problem. 
Let us first introduce this problem. 

\begin{Ex}
Let $S$ be a simple system such that:
\begin{enumerate}
\item There are two users, \alice and \bob, such that \alice (for instance, a physician) is more qualified than \bob (for instance, a nurse);
\item There are two resources, \high and \low, such that the resource \high (for instance, one of \alice's patient record) is more sensitive than \low (for instance, some indicates related to a drug);
\item \bob is not normally qualified enough to access \high;
\item in case of emergency (for instance, the patient is having a heart attack), the resource \high should be accessed. 
\end{enumerate}  
\end{Ex}
The role of the security engineer is therefore to define a mechanism that, given some the current state and an access $(u, r)$, where $u$ is a user and $r$ a resource, returns a security decision, such as \allow 
or \deny. The traditional approach to do so is to define a security policy, that is, a set of rules describing what decision to make for each access, and then to define a simple program that checks if the
given access satisfies the policy or not. For instance, we could define the following policy, for any access $(u, r)$:
\begin{quote}
  If $u = \bob$, $r = \high$ and there is no emergency, then \deny else \allow. 
\end{quote}
There exist many languages (e.g., XACML) in which this policy can be defined. However, 
defining a ``static'' policy requires to resolve beforehand all possible situations. In our example, 
there is a clear conflict between rules 3 and 4: in case of emergency, if \alice is not accessing \high, then either we allow \bob
to access it, thus breaking rule 3, or we do not, thus breaking rule 4. Note that there exist some approaches to deal with this particular problem 
of delegation (in case of unavailability, \alice should automatically delegate her right over \high to \bob), such as~\cite{cram:stm10,KMMY11}, 
however we want to illustrate a more general problem: the combination of rules might lead to conflicting situations. The resolution of such 
conflicts is usually addressed by considering composition operators, thus allowing some rules to take precedence over others. We propose
here a different approach, where conflict resolution is obtained as the result of an optimization process.

Indeed, we defined in~\cite{MM12} a novel approach, where the security policy is not defined, but {\em derived} from a decision process. Intuitively, the responsibility of the security engineer
is only to put some values on the different resource and/or states of the system, to describe the probabilistic behaviour of the system, and the optimal policy can be automatically defined. 
More precisely, the security engineer must defined an Access Control Markov Decision Process (AC-MDP), given by:

\begin{Def}
An AC-MDP is a tuple $\langle \Sigma, \A, \P, \W \rangle$, where: 
\begin{itemize}
\item $\Sigma = \I \times \R$ is a set of access control states, such that $\I$  represents the security information of the state and $\R$ the access requests (each state containing a request to control), 
\item $\A$ is a set of decisions, 
\item $\P : \Sigma \times \A \times \Sigma \to [0,1] $ is the probability function, such that $\P(\sigma_i, a, \sigma_j)$, which, for the sake of exposition, we also write $p^{a}_{ij}$, stands for the probability of reaching the state $\sigma_j$ by executing the decision $a$ from the state $\sigma_i$,
\item  $\W : \Sigma \times \A \times \Sigma \to \U$ is the reward function, such that $\W(\sigma_i, a, \sigma_j)$, also written as $w^a_{ij}$, stands for the reward associated with executing the decision $a$ from the state $\sigma_i$ and arriving in the state $\sigma_j$. 
\end{itemize}
\end{Def}

It is worth noting that we consider here that the reward is a {\em gain} function, and that therefore the objective of the process is to eventually {\em maximize} the accumulated rewards. We could equivalently 
consider the reward as a {\em cost}, and in this case the objective would be to minimize it. In the following, for the sake of clarity, we associate gain with positive rewards, and cost with negative rewards, 
the important point being to {\em compare} two values between each other. 

We write $q^a_i$ for the immediate reward for executing the decision $a$ in the state $\sigma_i$: 
\begin{equation}
\label{eq:immediate_reward}
q^a_i = \sum_{j} p^a_{ij} \cdot w^a_{ij}
\end{equation}
In this context, a security policy is a function $\delta:\Sigma \to \A$, and the value $V^\delta$ of each state for this policy can be defined as:
\begin{equation}
V^\delta(\sigma_i) = q^{\delta(\sigma_i)}_i + \beta \sum_j p^{\delta(\sigma_i)}_{ij} V^\delta(\sigma_j) \label{eq:value_mdp}
\end{equation}
where $0 \leqslant \beta \leqslant 1$ is a discount factor, indicating how much weight is put on the value of future states. 
The optimal policy $\delta^*$ is the policy that maximizes the value function, that is, for any policy $\delta$ and any state
$\sigma$, we have $V^{\delta^*}(\sigma) \geqslant V^\delta(\sigma)$. More formally, given a state $\sigma_i$, the optimal policy $\delta^*$
is given by:
\begin{equation}
\label{eq:optimal_simple}
\delta^{*}(\sigma_i) =\arg \max_{a \in \A} \,\, [q^{a}_{i} + \beta \sum_j p^a_{ij} \, V^{\delta^*}(\sigma_j)] 
\end{equation}
Clearly, this policy is not necessarily unique. For instance, in the degenerate case where all decisions are associated with the same reward, then any policy is optimal. This case would be equivalent to defining a policy 
for an access that has the same impact whether it is allowed or denied. In order to define the actual policy and to solve the non-determinism of choosing between several optimal policies, we could introduce an arbitrary
ordering among decisions, such that if in a given state, two decisions maximize the value, we take the smallest.

\medskip
With our approach, instead of defining directly the policy, the security engineer instantiates the AC-MDP (and in particular the 
function $\P$ and $\W$), and the optimal policy can be calculated as an optimization problem. We present in the following sections
how to perform such instantiation for the concrete example.


\section{Definition of the AC-MDP in GMPL}
\label{sec:general}

GLPK\footnote{Open source software available at \texttt{http://www.gnu.org/software/glpk/}} is  a piece of software intended for solving large-scale linear programming problems, and it supports
the GMPL language for modeling problems, which is a subset of AMPL. We now show how to implement an AC-MDP in GMPL. 
For the sake of exposition, we first present the ``model'' part of the implementation, that is, the 
part that does not depend on concrete values for the parameters, and we define in the following section some concrete values for these parameters. 
This implementation is directly inspired from that provided by Vincent Conitzer in his lecture on Linear and Integer Programming\footnote{\texttt{http://www.cs.duke.edu/courses/spring08/cps296.2/}}.

\subsection{States and Actions}

We first define the set $\I$ of security information and the set $\R$ of requests. 
Based on the problem description, we assume the qualifications of the entities to be fixed, thus not belonging to the state\footnote{It is of course possible to include qualification functions in the state, but that would make its structure more complex, with no real gain for the concrete example.}, and therefore we define the security information 
as a pair $(e, c)$, where $e$ is a boolean indicating whether there is a current emergency and $c$ is the set of current accesses, such that an access is simply a pair $(u, r)$. 
Furthermore, we model a request directly as an access. 

For the sake of generality, we define as parameters the number of users $\gm{NU}$ and resources $\gm{NR}$, and we define the set of users and resources as indices. 
We also introduce the ``empty'' request \gm{(eps,eps)}, that represents the fact that there is no access request to control. 

\lstset{frame=single}
\begin{gmpl}
param NU default 2;
set USERS := 0..(NU-1);

param NR default 2;
set RESOURCES := 0..(NR-1);

set ACCESS := USERS cross RESOURCES;

param eps, symbolic;
set FACCESS := ACCESS union {(eps, eps)};
\end{gmpl}

The state should include all previous currently granted accesses, however GMPL does not natively support powersets.  
Instead, we use a workaround by indexing each element of the powerset: we first associate each access $(u, r)$ with the 
index $2^{u * \gm{NR} + r}$ (e.g., in our settings, the index of $(0,0)$ is $2^0$, the index of $(1, 1)$ is $2^{1*2 + 1} = 2^3$, etc);
we then create the set \gm{PA}, which ranges from $0$ to $2^{\gm{NU*NR}}-1$; finally, we define the set \gm{POW} which is indexed
by \gm{PA}, such that, intuitively speaking, in the binary representation of \gm{k}, the $i$-th bit is equal to 1 if, and only if, 
\gm{POW[k]} contains the access indexed by $i$. For instance, for the index $\gm{k} = 5$, that is $\gm{k} = 101$ in binary, we have 
that \gm{POW[5]} contains exactly the accesses indexed by $2^0$ and $2^2$, i.e., the accesses $(0, 0)$ and $(1, 0)$. 
This encoding is defined in GMPL as follows. 
\begin{gmpl}
param n := (NU)*(NR);
set PA := 0 .. (2**n - 1);
set POW {k in PA} := setof{(u,r) in ACCESS:  (k div 2**(u * NR + r)) mod 2 = 1}(u,r);
\end{gmpl}
Note that the indexation of \gm{POW} is done over \gm{ACCESS} and not \gm{FACCESS}, meaning that, by construction, the empty request cannot belong to 
the set of previously granted accesses. A state is then given by an emergency status, a set of accesses and an access to control:
\begin{gmpl}
param calm, symbolic;
param alert, symbolic;
set EMERGENCY := {calm, alert};

set STATES := EMERGENCY cross PA cross FACCESS;
\end{gmpl}
For instance, the state \gm{[calm, 1, 1, 0]} represents the state of the system that is in a \gm{calm} status, where the access \gm{[0,0]} (corresponding to \gm{POW[1]}) has been previously 
granted and where the current access to control is \gm{[1,0]}\footnote{Note that in GMPL, there is an implicit ``inlining'' of the cartesian product, that is, in the above example, we do not write the 
state as \gm{[calm, 1, [1, 0]]}, even though the access \gm{[1, 0]} is a pair itself.}. 
Finally, we define the set of actions $\A = \{\allow, \deny\}$. 
\begin{gmpl}
param allow, symbolic;
param deny, symbolic;

set ACTIONS := {allow, deny};
\end{gmpl}

\subsection{Transition function}
\label{sec:transition-function}

We now show how to define the transition function $\P$, using three sub-functions, 
one for each component of a state. The probability of switching from one emergency 
status to another is given as a concrete parameter (we actually study the behaviour
of the model when this parameter is variating in Section~\ref{sec:emvar}) and 
therefore only declared in the model. 

\begin{gmpl}
param transition_emergency {e1 in EMERGENCY, e2 in EMERGENCY};
\end{gmpl}

Concerning the set of current accesses, for the sake of simplicity, we do not consider here
probabilistic modifications. In other words, whenever an access is allowed, the
following state contains it. However, it would be straightforward to extend this transition 
function to also consider the possibility that an access granted is not added, or even that 
a denied access is in fact performed. We could also model here the fact that at any time, there is 
a risk of leakage of information, that is, there is a non-null probability that a non requested 
access will be added to the set of current accesses in the next state. 

\begin{gmpl}
param transition_access {s1 in PA, (u, r) in FACCESS, a in ACTIONS, s2 in PA} := 
  if (a = deny or u = eps) then (if s1 = s2 then 1 else 0)
  else (if  ((POW[s2] within (POW[s1] union {(u, r)})) 
        and ((POW[s1] union {(u,r)}) within POW[s2])) then 1 else 0);
\end{gmpl}

An important point of our modelling is the fact that the state contains the request to control, 
and it follows that when defining the transition to the next state, we must also define what
will be the next request to control. We first assume that it is always \texttt{[eps, eps]} (i.e., we only control one request at the time), 
and we release this assumption in Section~\ref{sec:complexreq}. 

\begin{gmpl}
param transition_req {s1 in PA, (u,r) in FACCESS, a in ACTIONS, (u2,r2) in FACCESS} := 
      if (u2 = eps and r2 = eps) then 1 else 0;
\end{gmpl}

Finally, the transition probability is given by the product of the three previous sub-functions. 
\begin{gmpl}
param transition {(e1, s1, u1, r1) in STATES, a in ACTIONS, (e2, s2, u2, r2) in STATES} := 
      transition_emergency[e1, e2] * transition_access[s1, u1, r1, a, s2]
      * transition_req[s1, u1, r1, a, u2, r2];
\end{gmpl}

\subsection{Reward Function}

To some extent, defining the reward function corresponds to defining the policy, in the sense that 
instead of stating if an access is correct or not, the security engineer attaches a value to it, and this 
value will be later use to determine whether the access should be allowed or not. Hence, we declare to basic 
reward functions: \gm{reward\_access[u,r]} indicates the value gained by granting the access \gm{[u,r]}, 
and \gm{reward\_resources[r]} indicates the value of not accessing the resource \gm{r} in case 
of emergency. We instantiate these functions in Section~\ref{sec:concrete}. 
\begin{gmpl}
param reward_access {(u, r) in ACCESS};
param reward_resource {r in RESOURCES};
\end{gmpl}
From the function \gm{reward\_resource}, we can define the reward associated with a state, which is, 
in case of emergency, the sum of the reward of each resource. 
\begin{gmpl}
param reward_emresource {(e, s, u, r) in STATES} := 
  if (e = calm) then 0 else 
  sum {r1 in RESOURCES: forall{u1 in USERS} (u1,r1) not in POW[s]}  reward_resource[r1];
\end{gmpl}

Finally, we can define the reward associated with an entire transition, thus corresponding to the 
function $\W$ of the AC-MDP. 
\begin{gmpl}
param reward_transition {(e, s, u, r) in STATES, a in ACTIONS, (e2, s2, u2, r2) in STATES} := 
   (if u = eps and r = eps then 0) else 
   (if a = allow then reward_access[u,r] else 0) + (reward_emresource[e2,s2,u2,r2]);
\end{gmpl}
It is worth noting that if the state contains the empty request to control, then regardless of the decision, 
the reward of the transition is null.

\subsection{Value Function and Policy}
We can now define the optimization problem, in order to calculate the maximal value function, and thus the optimal policy. 
We first introduce the immediate reward, as defined in Equation~\eqref{eq:immediate_reward}. 
\begin{gmpl}
param immediate_reward {(e, s, u, r) in STATES, a in ACTIONS} := 
  sum {(e2, s2, u2, r2) in STATES} (transition[e,s,u,r,a,e2,s2,u2,r2] 
                                    * reward_transition[e,s,u,r,a,e2,s2,u2,r2]);
\end{gmpl}

Intuitively, we want to define the value function of the optimal policy $V^{\delta^*}$, that is, for each state 
$\sigma_i$, we want to define the equation: 
\[
V^{\delta^*}(\sigma_i) = \max_{a \in \A} [q^a_i + \beta \sum_j p^{a}_{ij} V^{\delta^*}(\sigma_j)] \label{eq:value_mdp}
\]
in which case we would just need to solve the corresponding system of equations. However, we cannot directly 
follow this approach, because the operator $\max$ is not linear. As stated by Conitzer in his lecture\footnote{\texttt{http://www.cs.duke.edu/courses/spring08/cps296.2/applications.pdf}}, the typical 
solution for this kind of problem is to define the equations such that $V^{\delta^*}(\sigma_i)$ must be greater or equal to  $q^a_i + \beta \sum_j p^{a}_{ij} V^{\delta^*}(\sigma_j)$, for any decision $a$, 
and to minimize the value of $V^{\delta^*}(\sigma_i)$, which thus corresponds to the lowest greater bound. In order to minimize each $V^{\delta^*}(\sigma_i)$, we simply aim at minimizing the sum 
of all the values. 

\begin{gmpl}
var value{(e,s,u,r) in STATES};

minimize total: sum{(e,s,u,r) in STATES} value[e,s,u,r];
s.t. bellman{(e,s,u,r) in STATES, a in ACTIONS}: value[e,s,u,r] >= 
   immediate_reward[e,s,u,r,a] 
 + sum{(e2, s2, u2, r2) in STATES}(beta*transition[e,s,u,r,a,e2,s2,u2,r2]*value[e2,s2,u2,r2]);
\end{gmpl}

\section{Instantiation}
\label{sec:concrete}

The model presented in the previous section represents a ``general'' implementation of the 
problem the security engineer wants to address: the number of users and resources can easily be 
extended, the rewards for accesses and states are not precisely defined, the probability of 
changing the emergency status is not specified, neither is the discount factor $\beta$. We show in this section 
how to define these particular values, and we describe the corresponding results. 

In order to compare the different results obtained, we focus on the value, from the empty state (i.e., 
where no access has been previously granted), of allowing or denying each access. More precisely, 
given an emergency status $e$ and an access $(u, r)$ such that $\sigma_i = (e, \emptyset, (u, r))$, and 
a decision $a$, we define the decision value $DV$ as:
\[
DV(e, (u,r), a) = q^{a}_{i} + \beta \sum_j p^a_{ij} \, V^{\delta^*}(\sigma_j)
\]
The security mechanism then selects the decision with the highest value.

\subsection{Basic Instantiation}
We first define a ``naive'' instantiation, somewhat equivalent to the static policy  
presented in Section~\ref{sec:problem}. In order to remove the values of the future 
states from the equation, we set the discount factor to 0. We also assume that the emergency status 
cannot change (i.e. we should make the decision considering that the status will not change). 
\begin{gmpl}
param transition_emergency:=  [*, *]:      
       calm   alert	:=
calm   1      0
alert  0      1;
\end{gmpl}

We also give arbitrary values for the reward function, trying to keep the idea of the original policy (for the sake 
of readability, we have used the names of the users and resources instead of their indices, as it is done in the actual 
implementation). 

\begin{gmpl}
param reward_access :=   [*,*]:
       high    low   :=
alice  6       10
bob    4       -10;
 
param reward_resource :=  low  0  high -20;
\end{gmpl}
As for the policy defined in Section~\ref{sec:problem}, we consider here that the reward for leaving the resource \high non accessed
in case of emergency is worse than letting \bob accessing \high. However, as we will see in the following, this does not necessarily 
imply that \bob will always be able to access \high. 

Since the discount factor $\beta$ is set to $0$, the decision value is directly equal to the immediate reward.
For instance, if we allow the access $(\alice, \low)$, we get an immediate reward of 6, while allowing the access 
$(\bob, \high)$ brings a reward of -10. Furthermore, if the reached state is in an emergency status, and the 
resource $\high$ is not accessed, we also get a reward of -20.

\begin{table}[t]
  \centering
  \caption{Decision values for $\beta = 0$ and emergency probability of $0$}
  \label{tab:basicvalues}
\begin{tabular}{|c|c|c|c|c|c|} \hline
 Status & Decision & (\alice, \low) & (\alice, \high) & (\bob, \low) & (\bob, \high) \\ \hline \hline
\multirow{2}{*}{Calm}    & \deny    & 0              & 0               & 0            & 0 \\
   & \allow   & 6              & 10              & 4            & -10 \\ \hline
\multirow{2}{*}{Alert}   & \deny    & -20            & -20             & -20          & -20 \\
  & \allow   & -14            & 10              & -16          & -10 \\ \hline 
\end{tabular}  
\end{table}
We present in Table~\ref{tab:basicvalues} the decision values for these 
basic parameters. The only situation where the decision value 
of \deny is higher than that of \allow is when \bob wants to access \high in a \gm{calm} status. It is however worth noticing that when there is an 
alert, both decision values for the access $(\alice, \low)$ are negative, meaning that both decisions are ``bad'' (since \high will not be accessed 
whatever decision is made), but the decision \allow is still better than \deny. This is an interesting aspect of quantitative approaches, as they 
enable the system to make the ``best'' decision, even though this decision might lead to a ``bad'' situation. 

\subsection{Complex Requests}
\label{sec:complexreq}

The previous instantiation considers a very ``static'' configuration of the system, since the emergency status is not assumed to dynamically change 
and the discount factor is null. In practice, a system is not frozen in time, and does not stop after controlling one request. Hence, in order 
to make the ``best'' decision, it is necessary to take into account the possible future evolutions of the system. In the following, in addition 
to setting the discount factor to a non null value, we also consider two parameter variations: there is a non-null probability that a state 
in a calm status will transition to a state in an alert status, and another request might need to be controlled after the current one. 
In order to specify the former possibility, we define:
\begin{gmpl}
param transition_emergency:=  [*, *]:
       calm	alert	:=
calm   0.9      0.1
alert  0        1;
\end{gmpl}
We then characterize the set of requests that are possible after executing a given action over 
a given request from a given set of accesses. More specifically, we distinguish between three cases\footnote{For the sake of readability, we skip the implementation details. All the different GMPL programs presented here can be found at \texttt{http://www.morisset.eu/qasa12/}}: 
\begin{compactitem}
\item \unique: only one request is controlled, meaning that the only possible following request is the empty request. 
\item \once: each request can only be allowed once. 
\item \all: all non-empty requests are possible (simulating an infinite trace): 
\end{compactitem}

We present in Table~\ref{tab:complexvalues} the decision values obtained by using the probability of change of the 
emergency status from \gm{calm} to \gm{alert} to $1$.

\begin{table}[t]
\centering
  \caption{Decision values for $\beta = 0.9$ and emergency probability of $0.1$}
  \label{tab:complexvalues}
\begin{tabular}{|c|c|c|c|c|c|c|} \hline
 ~Request~ & ~Status~ & ~Decision~ & (\alice, \low) & (\alice, \high) & (\bob, \low) & (\bob, \high) \\ \hline \hline
\multirow{4}{*}{\unique} & \multirow{2}{*}{Calm}    & \deny    & -2              & -2               & -2            & -2 \\
&    & \allow   & 4              & 10              & 2            & -10 \\ \cline{2-7}
& \multirow{2}{*}{Alert}   & \deny    & -20            & -20             & -20          & -20 \\
&   & \allow   & -14            & 10              & -16          & -10 \\ \hline \hline

\multirow{4}{*}{\once} & \multirow{2}{*}{Calm}    & \deny    & 2.63              & 2.63               & 2.63            & 2.63 \\
&    & \allow   & 7.58              & 14.15              & 6.41            & -1.59 \\ \cline{2-7}
& \multirow{2}{*}{Alert}   & \deny    & -23.54            & -23.54              & -23.54           & -23.54  \\
&   & \allow   & -15.54            & 14.15              & -16.70          & -1.59 \\ \hline \hline

\multirow{4}{*}{\all} & \multirow{2}{*}{Calm}    & \deny    & 34.80              & 34.80               & 34.80            & 34.80 \\
&    & \allow   & 40.80             & 55              & 38.80            & 35 \\ \cline{2-7}
& \multirow{2}{*}{Alert}   & \deny    & 4.55            & 4.55             & 4.55          & 4.55 \\
&   & \allow   & 10.55            & 55              & 8.55          & 35 \\ \hline 
\end{tabular}  

\end{table}

It is worth observing that the request transition has an important impact on the actual decision values, but the ordering is almost always
preserved. Indeed, the decision value of \allow is always higher than that of \deny for the accesses $(\alice, \low)$, $(\alice, \high)$ 
and $(\bob, \low)$, and for the access $(\bob, \high)$ when the state is known to be in the \gm{alert} status. However, from a state 
in a \gm{calm} status, for the behaviours \once and \unique, it is better to deny the access $(\bob, \high)$ while for the 
behaviour \all, it is better to allow it. 

This result is somewhat counter-intuitive, since one could expect that for \once and \unique, if we deny the access, the probability that 
the resource \high will be accessed is quite low, while for \all, even if we deny the access to \bob, we know that \alice will ask for the 
access in the future, and therefore that \high will be eventually accessed. In fact, this result is due to the fact that we automatically 
associate the state containing the empty request with a null value. In other words, we do not take into account the fact that the 
system might stay forever in a state where \high is not accessed. On the other hand, for \all, we have somehow infinite sequences of 
requests, which automatically gives a high value to all states. 

We can modify the behaviour such that starting from state a containing the empty request takes the reward of the reached state (i.e., in our 
example, the same state) into account:

\begin{gmpl}
param reward_transition {(e, s, u, r) in STATES, a in ACTIONS, (e2, s2, u2, r2) in STATES} := 
  (if a = allow and u != eps then reward_access[u,r] else 0) 
   + (reward_emresource[e2,s2,u2,r2]);
\end{gmpl}

In this case, the decision values for the \all behaviour do not change, on the contrary of the \unique and \once 
behaviours, for which the ordering of the decisions for the access $(\bob, \high)$ in the \gm{calm} status actually change. 
Indeed, for \unique, denying this access leads to a value of -105.26, while allowing it leads to -10. 
Similarly, for \once, denying it leads to a value of -32.35, while allowing it leads to -1.59. In other words, the values
for allowing the accesses are identical to those in Table~\ref{tab:complexvalues}, but the values for denying are much lower, 
thus reflecting the idea that the state is likely to remain in a ``bad'' state, where \high is not accessed. 

Nonetheless, it is interesting to observe that although the ordering between the different behaviors is usually the same,
meaning that we make decisions consistent with the original policy, the difference between the value of each decisions can significantly vary. 
For instance, consider the access $(\alice, \low)$ for the behaviour \all in the \gm{calm} status: denying it leads to a value
of 34.80, while allowing it ``only'' leads to a value of 40.80. In other words, the gain of allowing versus denying it is quite low, relatively and
absolutely speaking. If we were to attribute a priority for the treatment of this access, we could probably give it a low one. 

On the other hand, for \all, from the \gm{alert} status, the differential for the access $(\alice, \high)$ is very important: the value of 
denying it is 4.55, while allowing it brings a value of 55, i.e., multiplying the value by a factor greater than 10. In other words, there is no 
doubt at all that this access should be allowed in these conditions. It is therefore interesting to observe that in addition to the decision, 
our approach also provides the ``strength'' or ``confidence'' in the decision. 


\begin{figure}[htp] \centering

\subfigure[Decision values for \unique,\high]{
\includegraphics[width=0.47\linewidth, trim=60px 60px 60px 60px]{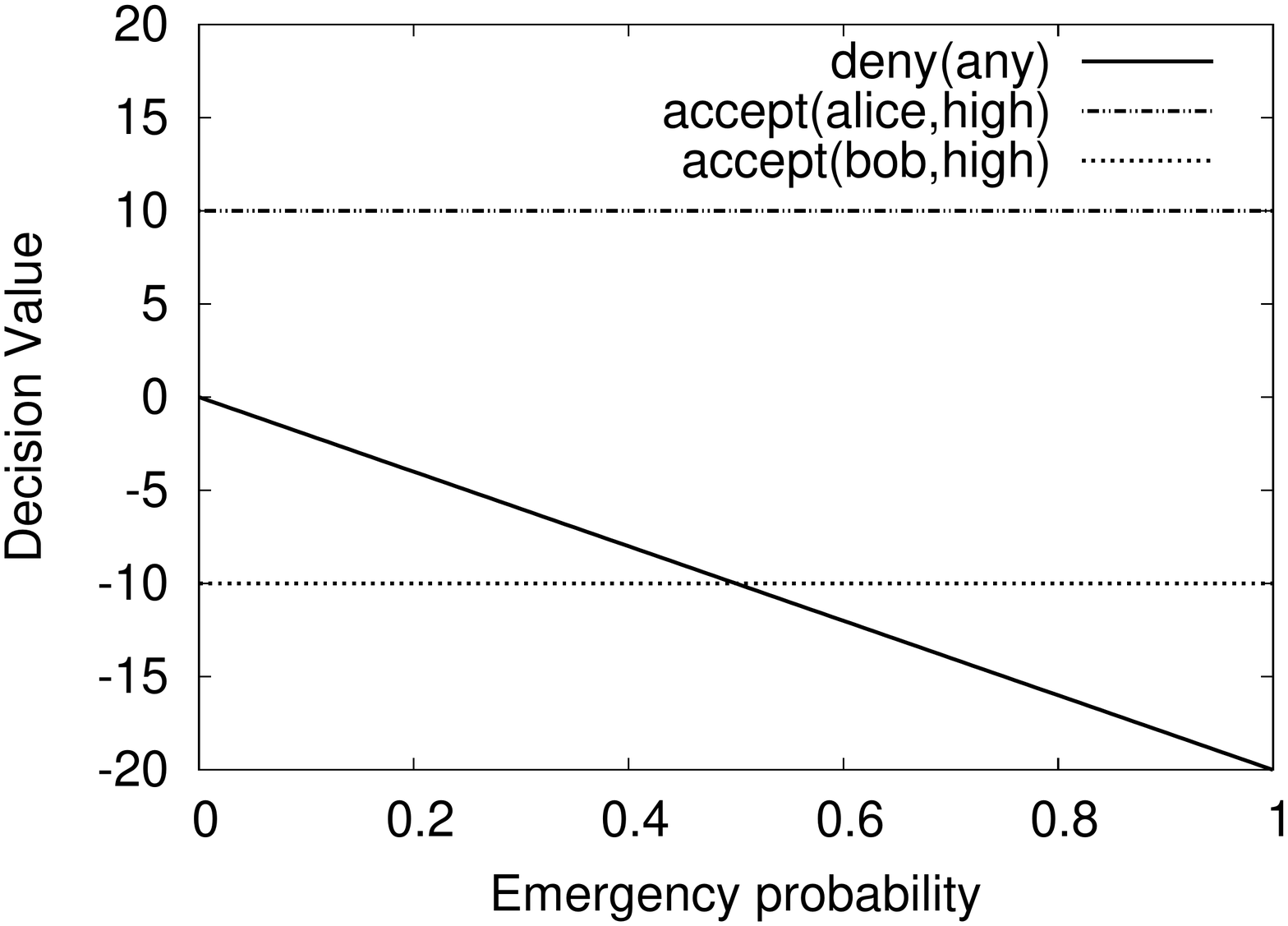}
}
\subfigure[Decision values for \unique,\low]{
\includegraphics[width=0.47\linewidth, trim=60px 60px 60px 60px]{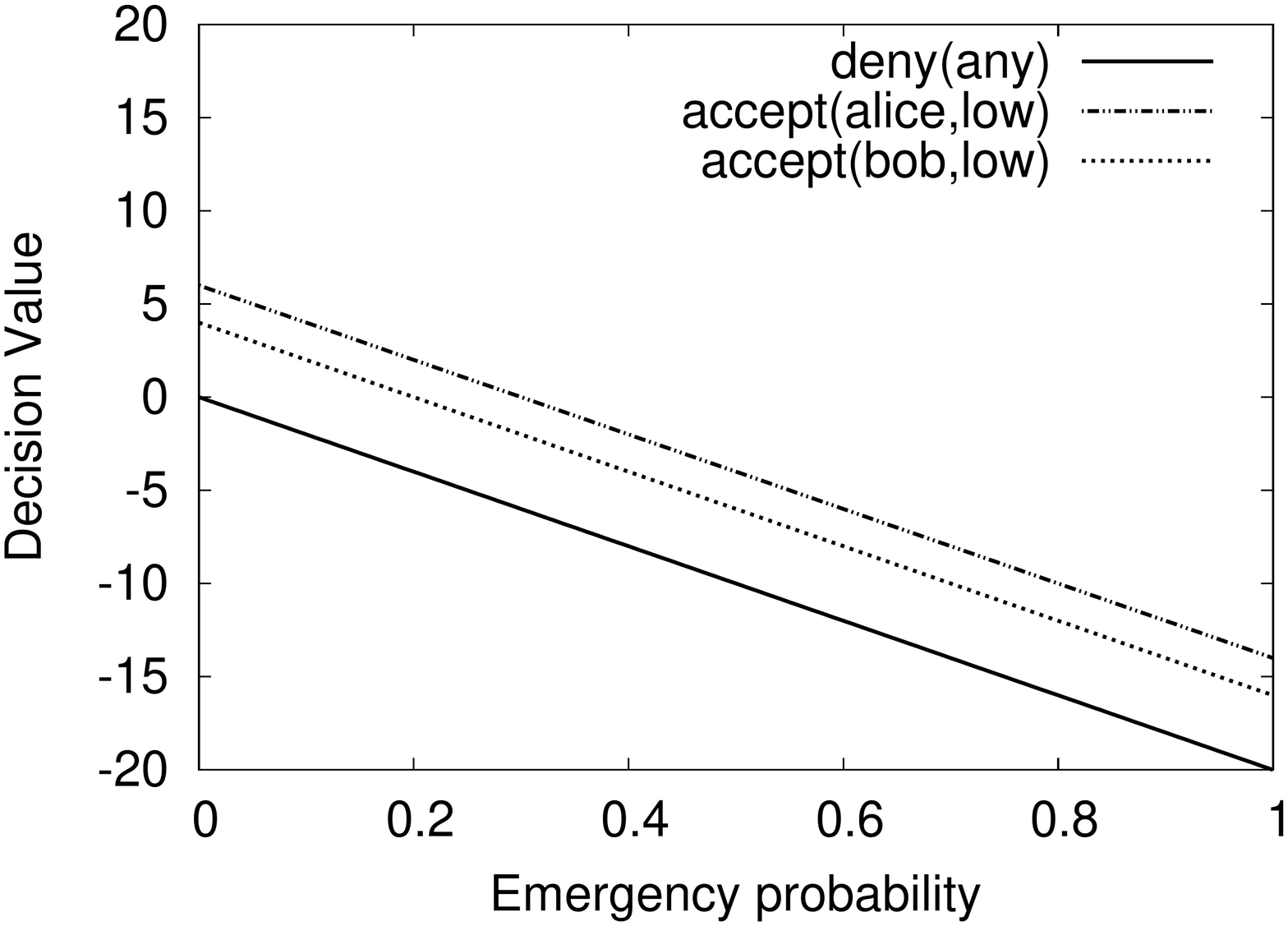}
}

\subfigure[Decision values for \once,\high]{
\includegraphics[width=0.47\linewidth, trim=60px 60px 60px 60px]{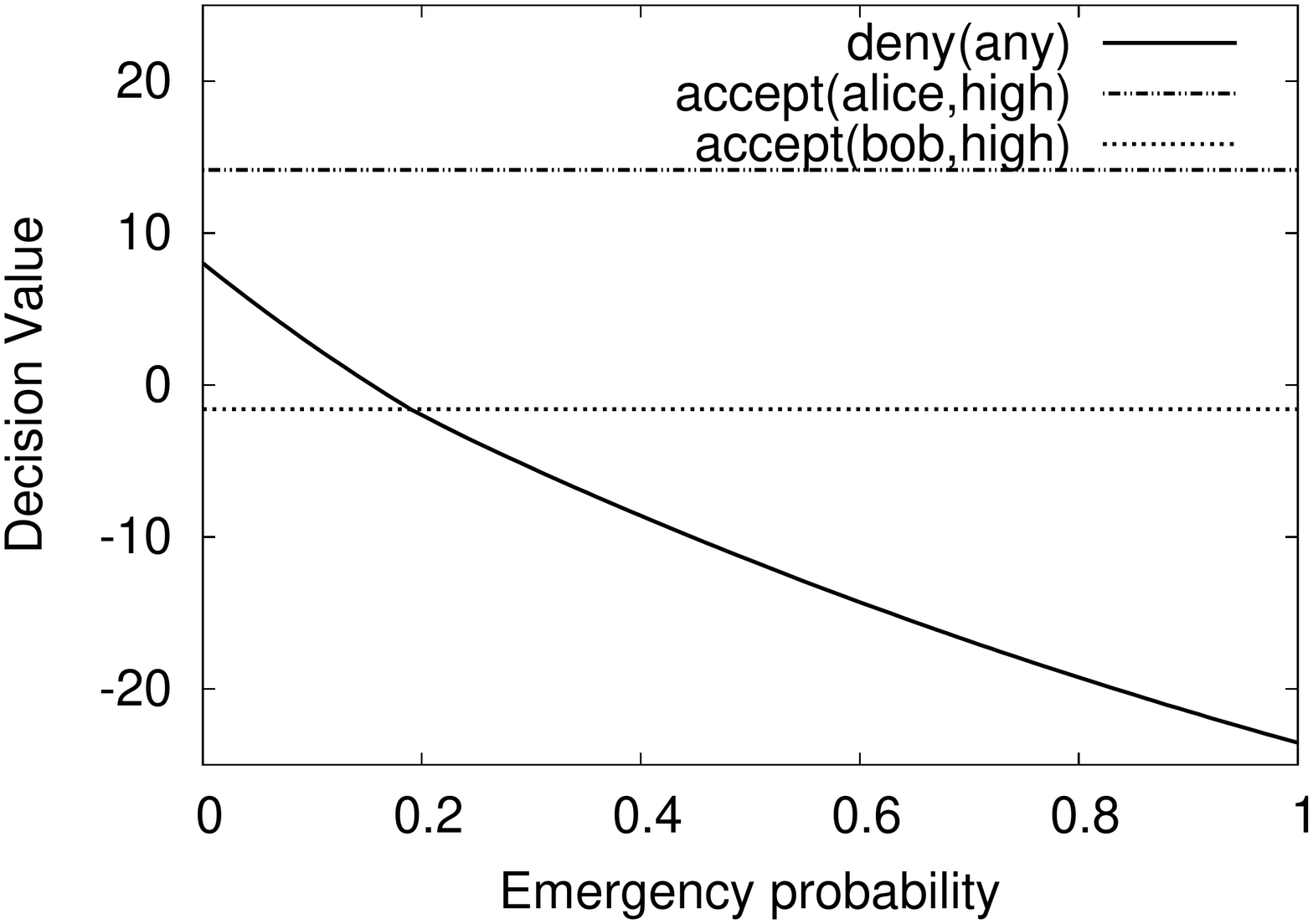}
}
\subfigure[Decision values for \once,\low]{
\includegraphics[width=0.47\linewidth, trim=60px 60px 60px 60px]{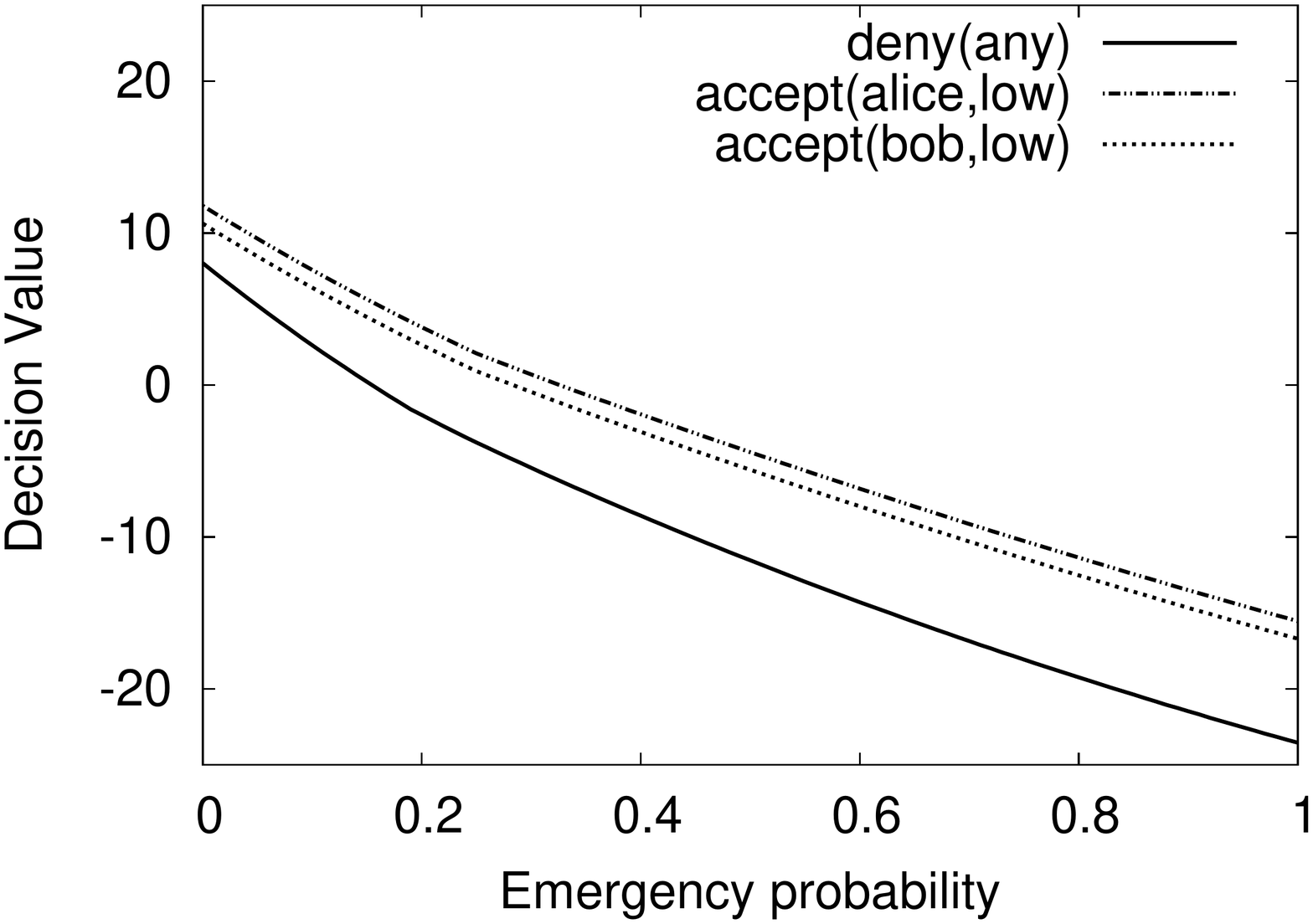}
}

\subfigure[Decision values for \all,\high]{
\includegraphics[width=0.47\linewidth, trim=60px 60px 60px 60px]{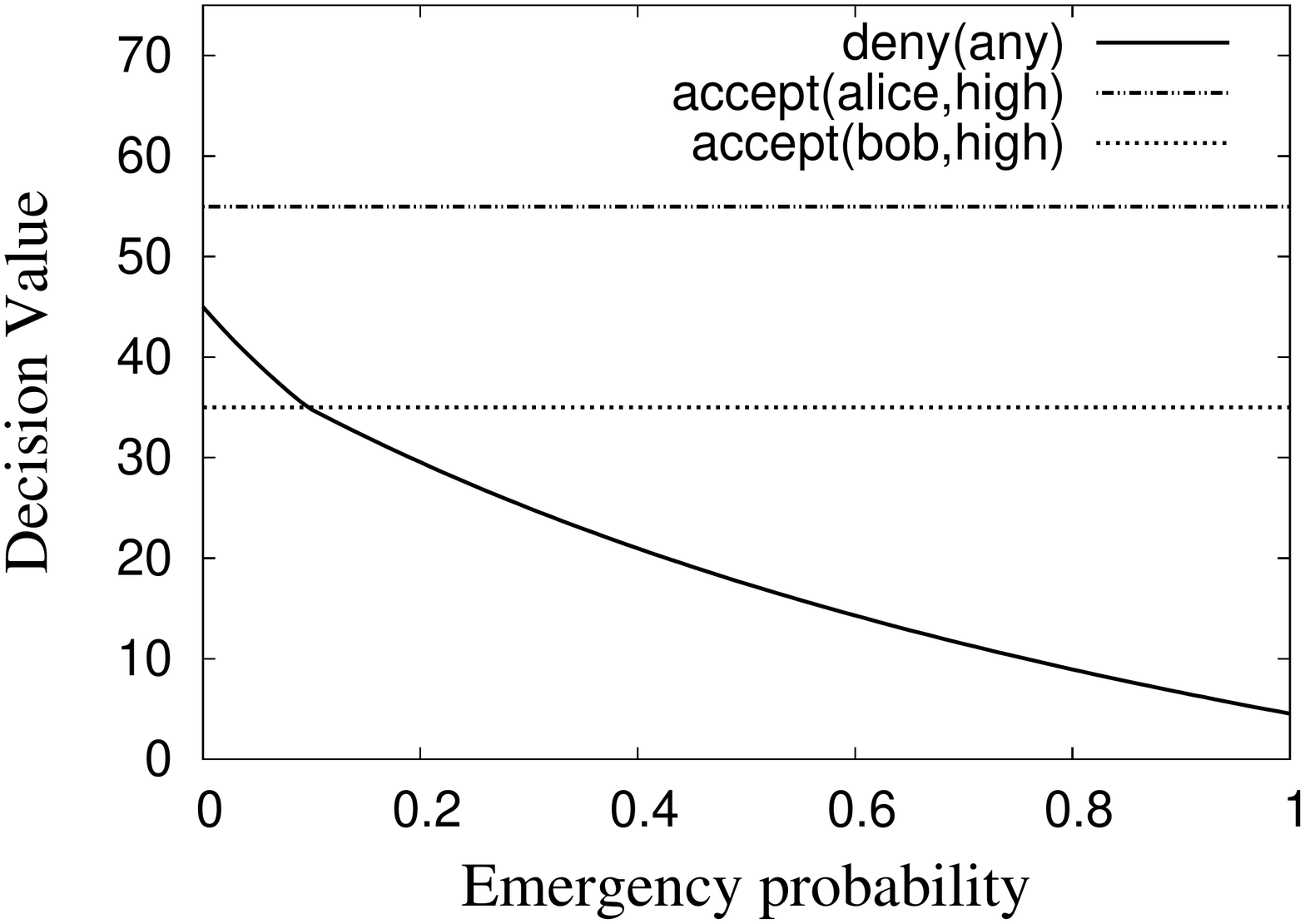}
}
\subfigure[Decision values for \all,\low]{
\includegraphics[width=0.47\linewidth, trim=60px 60px 60px 60px]{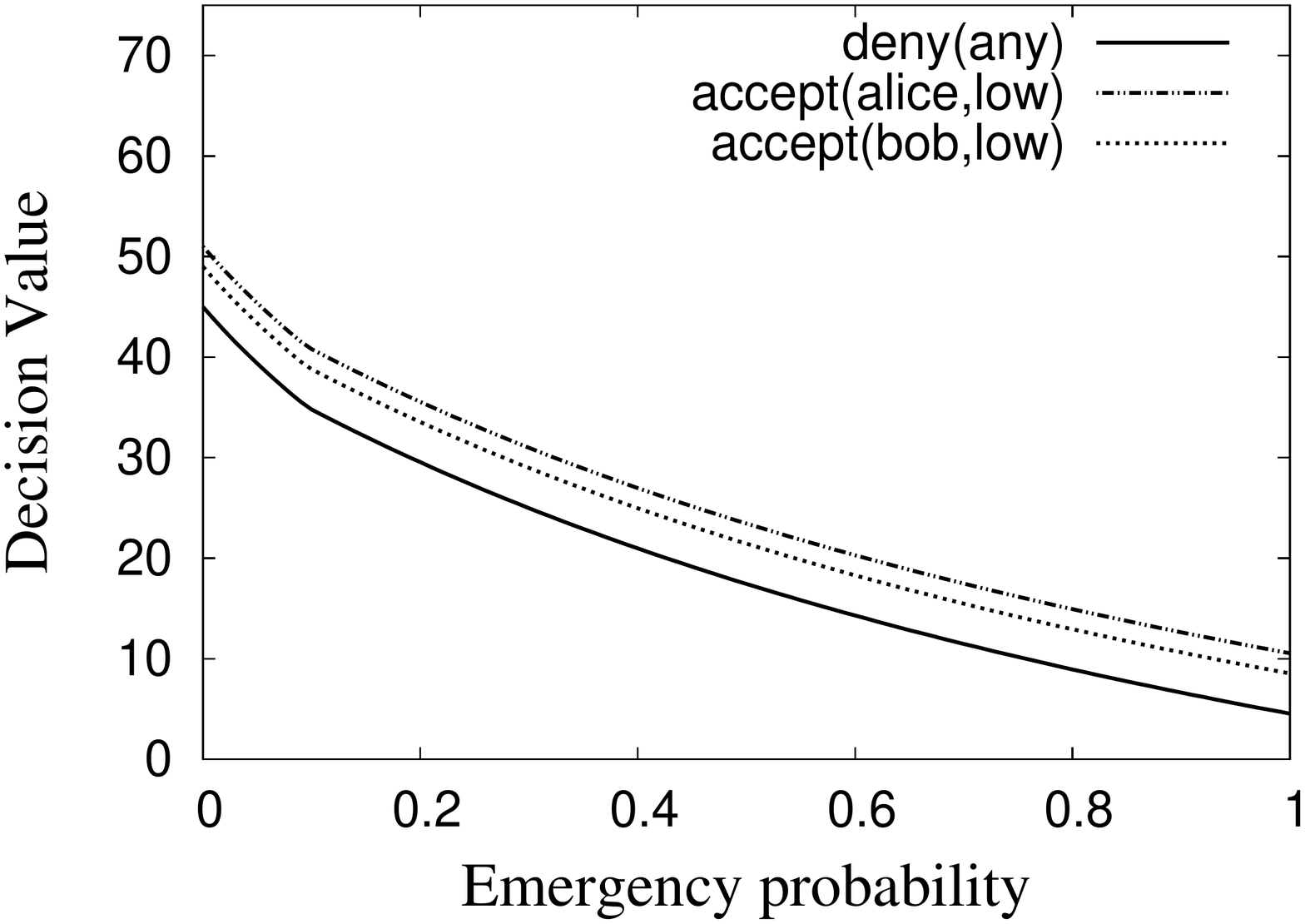}
}
\caption{Decision values for $\beta = 0.9$, where the emergency probability stands 
for \gm{transition_emergency[calm,alert]}}
\label{fig:variation}
\end{figure}

\subsection{Variation of the emergency probability}
\label{sec:emvar}

Finally, our last experiment consists in calculating how, for a given discount factor ($\beta = 0.9)$, the decision values vary when 
the emergency probability varies. Intuitively, if the system is very stable (i.e., the probability equals 0), then any transition 
from a calm state leads to a calm state, and the system does not need to care whether $\high$ is accessed or not. On the other hand, 
if the system is very unstable (i.e., the probability equals 1), then any transition will lead to an emergency state, in which case 
the system has to ensure as much as possible that the resource $\high$ is accessed. 

We show in Figure~\ref{fig:variation} these results, for each request behaviour. 
Firstly, we can observe that for the resource \low, the value of denying an access to it, regardless of the user, is always
lower than allowing it. It therefore fits with the intuition that whatever the emergency status, there is no gain in denying
an access to the resource \low. 
We can similarly observe that for a given resource, the value of allowing the access for \alice is always greater than 
allowing the access for \bob, which is consistent with the requirement that \alice is more qualified than \bob. 

Finally, it is interesting to see that the emergency probability required for the value of allowing $(\bob, \high)$ to be higher 
than that of denying it changes according to the request behaviour: it is exactly 0.5 for \unique, around 0.19 for \once and between 0.09 and 0.1 for \all. 
This observation means that in order to make the best decision, it is not enough to know the values for each access and the 
probability of emergency, one also needs to know what will be the future behaviour of the system and/or of the user.

\section{Discussion - Conclusion}
\label{sec:discussion}

In this paper, we started from a very concrete problem, loosely inspired from the healthcare context, and we gradually implemented 
a security mechanism based on an Access Control Markov Decision Process. We have shown how changing the different parameters can have
an impact on the final decision values. It is worth mentioning that the computational complexity does not change when the parameters change: in other
words, it takes the same time and the same memory to compute all decision values in Table~\ref{tab:basicvalues} than to compute those for a specific
request behaviour as in Table~\ref{tab:complexvalues}. Similarly, it could be possible to create more complex reward functions, the significant 
computational parameter being the number of states, not the way the reward function is calculated (although these two notions can of course be related, 
i.e., in order to have an expressive reward function, one might need to extend the state). 
We believe that this approach raises the following observations and questions: 

\noindent
$i)$ It is possible to define an access control mechanism taking into account different reward functions
using linear programming for two users and two resources. What about several dozens of users and several thousand of files? It is probably 
not reasonable to expect to find the optimal policy at runtime, but it could be done at compile time, since we actually compute the values for {\em all} states. 

\noindent
$ii)$ The ordering between the value of two decisions defines the policy (i.e., we pick the decision with the highest value), but the differential 
between these values is also significant, and could be used for instance to represent a notion of priority. Can we use this difference as an intuitive idea of 
trust? 

\noindent
$iii)$ The future behaviour of the system has an impact: the best decision for a single access might be different whether another access will be submitted 
in the future, and we can calculate what this difference will be, but how reasonable is it to predict what will be the future behaviour? 

\noindent
$iv)$ Our approach can also be used as an analysis tool, in order to understand the impact of changing one parameter, but how to define the initial parameters? How to know 
whether the reward $(\bob, \high)$ is -10 and not -42? 

Clearly, we probably bring more questions than answers, but we believe that using quantitative tools can help the definition of security mechanisms, in particular by 
providing the security engineer with a new way of thinking her security problem, and we hope to pave the way towards a generalized usage of such tools.


\end{document}